\documentclass[prl,amsmath,amssymb,superscriptaddress,twocolumn]{revtex4}
\usepackage{graphicx}

\newcommand{\br}{{\bf r}}

\newcommand{\bp}{{\bf p}}

\newcommand{\bq}{{\bf q}}

\newcommand{\eps}{\epsilon}

\DeclareMathAlphabet{\mathpzc}{OT1}{pzc}{m}{it} 
\begin{document}
\title{Anomalous thermodynamics of Coulomb interacting massless Dirac fermions in two spatial dimensions}
\author{Oskar Vafek}
\affiliation{National High Magnetic Field Laboratory and Department
of Physics,\\ Florida State University, Tallahassee, Florida 32306,
USA}
\date{\today}
\begin{abstract}
It is argued that the specific heat of $N$ massless Dirac fermions
in 2 spatial dimensions interacting with 1/r Coulomb interactions is
suppressed logarithmically relative to its non-interacting
counterpart. The (dimensionless) coefficient of the logarithm is
calculated in a closed form in the leading order in large $N$
expansion, but to all orders in the effective fine structure
constant, $\alpha_F$, a procedure which takes into account finite
temperature screening. This effect is expected to occur in a single
layer graphene embedded in a dielectric medium. Its dependence on
the dielectric constant is calculated analytically.
\end{abstract}
\maketitle

Understanding properties of quantum matter confined to two spatial
dimensions has been at the forefront of theoretical physics
\cite{Kivelson1996}. The lowered dimensionality is believed to
increase the role of interactions and, together with quantum
statistics, lead to anomalies in various physical observables
\cite{Bonn2006,Kravchenko2001,Casey2003}.

In the case of Fermi-Dirac systems, the role of (short range)
interactions has been studied in the pioneering work of
Landau\cite{Landau1957}. What is now understood as the Fermi liquid
paradigm explains the low temperature properties of the degenerate
Fermi-Dirac systems as arising from a dilute gas of weakly
interacting quasiparticles whose quantum numbers are the same as
those of non-interacting fermions. In this context, it is also well
known that a degenerate assembly of electrons with a rotationally
invariant Fermi surface, in either three or two spatial dimensions,
interacting with $1/r$ Coulomb interactions, can be described by the
Fermi liquid theory \cite{PinesNozieres}. Ultimately, this is due to
the ability of the mobile carriers to {\it screen} the long ranged
interaction beyond the distances of the order of the inter-electron
separation. One of the most direct observable consequences for a
metal is the linear temperature dependence of the specific heat at
asymptotically low temperatures i.e. $\lim_{T\rightarrow
0}c_V/T=\gamma$ \cite{GellMann1957,nonanalytic}.

The situation is less clear in the case of a semimetal, which can be
thought of loosely as the Fermi surface shrunk to a point. More
precisely, in this case, a conduction band and a valence band
typically touch at a discrete set of points with dispersion which
vanishes linearly with the wavenumber near each such point, being in
a sense a critical point between a metal and an insulator. If the
chemical potential lies at such a point of degeneracy, the absence
of the Fermi surface implies that, unlike in a metal, the long
ranged interactions are screened only by thermally excited carriers,
with the screening length of the order of the thermal length, $\hbar
v_F/k_BT$. For a typical Fermi velocity $\sim 10^6m/s$, even at room
temperature the screening length can therefore be much longer than
the interatomic spacing. The effect of the Coulomb interactions on
3D semimetals was investigated in Ref.
\cite{AbrikosovBeneslavskii1971} where it was argued that, unlike in
a metal, the Coulomb interaction causes the energy spectrum to
differ from a purely linear spectrum by a logarithmic factor.

In this work, I study the effects of the Coulomb interactions on the
thermodynamic properties of massless Dirac fermions in 2D, which can
be used as a low energy description of a single-layer graphene, and
find that the poorly screened electron-electron interaction leads to
anomalies in the quasiparticle thermodynamics.

In the absence of interactions, the free energy density is (up to a
temperature independent constant) $$
f_0=-Nk_BT^3\frac{3\zeta(3)}{4\pi v_F^2},
$$
where $N$ is the number of the 2 component Dirac flavors ($N=4$ in
the single layer graphene) and $\zeta(n)$ is the Riemann zeta
function. The strength of the electron-electron interaction is given
by the dimensionless coupling constant $\alpha_F=e^2/(\eps\hbar
v_F)$, which is the effective fine structure constant, where $\eps$
describes the polarizability of the surrounding medium. If $\eps$ is
large, then $\alpha_F$ is small, and to first order in $\alpha_F$,
$v_F$ receives a logarithmic correction $\frac{1}{4}v_F
\alpha_F\ln[\frac{\Lambda}{k}]$, where $\Lambda$ is a short
wavelength cuttoff. Such logarithmic enhancement of the otherwise
linear dispersion near the Dirac point, would imply logarithmic
suppression of the specific heat. It is not clear, however, whether
such simpleminded reasoning suffices. Similar reasoning in the case
of an electron gas with a Fermi surface would also lead to a
logarithmic suppression of the effective mass, the result known to
be incorrect due to screening. Therefore, the key issue in this
context is the effect of the {\em thermal} screening on the specific
heat in the 2D Dirac case.

To answer this question, I set up a large $N$ expansion while
keeping $\alpha_FN$ fixed and arbitrary. To the leading order in
large $N$, this is equivalent to a random phase approximation (RPA).
I find that the free energy density receives a logarithmic
correction
\begin{eqnarray}\label{deltaf1}
\delta
f=\frac{3\zeta(3)}{\pi^2}\left[\frac{8}{\pi}\lambda^2g(\lambda)-\frac{2}{\lambda}\right]\frac{k_B^3T^3}{\hbar^2v_F^2}
\ln\left(\frac{T_{UV}}{T}\right)
\end{eqnarray}
where $\lambda=\alpha_F N\pi/8$, the large temperature cutoff is
related to the short wavelength cutoff  by $\hbar
v_F\Lambda=k_BT_{UV}$, and the function $g$ is given by the
Eq.(\ref{g}).

\begin{figure}[t]
\begin{center}
\includegraphics[width=0.4\textwidth]{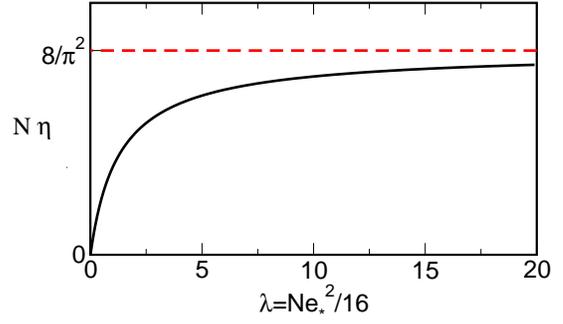}
\end{center}
\caption{The coefficient
$\eta=\frac{1}{N}\left[\frac{16}{\pi^2}\lambda^2g(\lambda)-\frac{4}{\pi\lambda}\right]$
of the logarithmic increase of the free energy in Eq.
(\ref{fansatz}) vs. $\lambda=Ne_{\ast}^2/16=N\frac{\pi}{8}
\frac{e^2}{\eps\hbar v_F}$. The closed form expression for $g$ is
given in Eq.(\ref{g}). For small $\lambda$, $\eta =2\lambda/(\pi N)$
while for $\lambda \rightarrow \infty $, $\eta=8/(\pi^2N)$.}
\label{etaPlot}
\end{figure}

To the same order in the large $N$ expansion, the above result can
be interpreted as a free energy density of $2N$ modes \cite{cone}
with dispersion
$$
\eps_{\eta}(k)=\hbar v_F
k\left(1+\eta\ln\left[\frac{\Lambda}{k}+1\right]\right)
$$
where
$$
\eta=\frac{1}{N}\left[\frac{16}{\pi^2}\lambda^2g(\lambda)-\frac{4}{\pi\lambda}\right].
$$
I find that as $\lambda\rightarrow 0$,  $\eta =2\lambda/(\pi N)$
while for $\lambda \rightarrow \infty $, $\eta=8/(\pi^2N)$ (see
Fig.\ref{etaPlot}).
 The expression (\ref{deltaf1}) can be understood as the first
non-trivial term in the Taylor expansion of
\begin{eqnarray}\label{fansatz}
f(T,\Lambda,\eta)=-2N k_BT \int_0^{\infty} \frac{dk}{2\pi}k
\ln\left[1+\exp\left(-\frac{\eps_{\eta}(k)}{k_BT}\right)\right].
\end{eqnarray}
Note that in the limit of $\lambda\rightarrow 0$, the above
expression coincides with the free energy calculated within the
Hartree-Fock approximation. As $T\rightarrow 0$, the free energy in
(\ref{fansatz}) vanishes as $-T^3/\ln^2T$\cite{logs,Corless1996}.
Thus the suppression of the specific heat
$c_V=-T\partial^2f/\partial T^2$ relative to the non-interacting
case {\em persists} when the polarization effects are included.

The logarithmic divergence of the particle's group velocity at low
wavenumbers may seem to violate causality. This is an artifact of
the approximation which treats the $1/r^2$ Coulomb forces as
instantaneous. Were we to include the retardation effects introduced
by coupling the fermions to a (quantum) 3D electromagnetic gauge
field, the Fermi velocity would not grow without bound at low $k$,
but instead saturate to the speed of light \cite{Gonzales1994}.
However, since the growth of $v_F$ is only logarithmic, the effects
of retardation on the specific heat would be practically
unobservable for $c\gg v_F$. For all practical purposes, the above
non-relativistic expression will therefore span the physically
relevant regime.

Below, I justify the above claims. For convenience, I work in units
where $\hbar=k_B=v_F=1$ and restore the physical units in the final
expressions. Then, the Hamiltonian is
\begin{eqnarray}\label{H}
\mathcal{H}=\mathcal{H}_0+\hat{V}.
\end{eqnarray}
The free part of $\mathcal{H}$ is
\begin{eqnarray}
\mathcal{H}_0=\sum^N_{j=1}\int d^2\br
\left[\psi_{j}^{\dagger}(\br)\bp\cdot\sigma\psi_j(\br)\right]
\end{eqnarray}
where $\bp=-i\nabla$, $\sigma$'s are the Pauli matrices, and
$\psi_{j}(\br)$'s are two component anticommuting Fermi fields. $N$
is the number of the Dirac fermion flavors which, in the condensed
matter setting, equals the number of valleys times the number of
spin directions ($N=4$ for a single-layer graphene). The interaction
part of $\mathcal{H}$ comes from the (3D) Coulomb interaction
between the charge density fluctuations and reads
\begin{eqnarray}
\hat{V}=\frac{1}{2}\int d^2\br d^2\br'\left[\delta
\hat{n}(\br)\frac{e^2}{\eps}\frac{1}{|\br-\br'|}\delta
\hat{n}(\br')\right].
\end{eqnarray}
Here
$\delta\hat{n}(\br)=\sum_{j}\psi_{j}^{\dagger}(\br)\psi_j(\br)$, and
$\eps$ is dielectric constant of any potential surrounding
insulator. Note that in our units $e^2/\eps$, which is the analog of
a fine structure constant, is dimensionless.

Under a scale transformation $\br \rightarrow b\br$ the Hamiltonian
(\ref{H}) appears to transform as $\mathcal{H}\rightarrow
b^{-1}\mathcal{H}$. This might suggest that even in the interacting
theory, the temperature dependent part of the free energy density
$f$ goes as $T^3$. However, since the theory is well defined in the
UV only in the presence a short distance cutoff $\Lambda^{-1}$, this
symmetry and the $T^3$ form will in general be violated.

In the large $N$ expansion the leading non-trivial correction to the
free energy density due to interactions is
\begin{eqnarray}\label{deltaf}
\delta f\!\!&=&\!\!\int_0^{\Lambda}\frac{dq
q}{2\pi}\int_{0}^{\infty}\!\!\frac{d\Omega}{2\pi}\coth\frac{\Omega}{2T}
\left\{\tan^{-1}\!\!\left[\frac{\Im
m\Pi^{ret}_0(q,\Omega,T)}{\frac{q}{e_{\ast}^2}+\Re
e\Pi^{ret}_0(q,\Omega,T)}\right]\right.\nonumber\\
&-&\left.\tan^{-1}\left[\frac{\Im
m\Pi^{ret}_0(q,\Omega,0)}{\frac{q}{e_{\ast}^2}+\Re
e\Pi^{ret}_0(q,\Omega,0)}\right] \right\},
\end{eqnarray}
where $e^2_{\ast}=2\pi e^2/\eps$. $\Pi^{ret}_0(q,\omega,T)$ is the
retarded polarization function of the free fermions.
\begin{widetext}
For $\Omega^2>\bq^2$:
\begin{eqnarray}\label{ReDyn}
\Re e\Pi_0^{ret}(q,\Omega,T)&=&NT\frac{\ln2}{\pi}-
\frac{Nq^2}{4\pi\sqrt{\Omega^2-q^2}}\int_1^{\infty}dx\left[\frac{\sqrt{x^2-1}}{1+\exp{\left|\frac{|\Omega|-qx}{2T}\right|}}
-\frac{\sqrt{x^2-1}}{1+\exp{\left|\frac{|\Omega|+qx}{2T}\right|}}
\right]
\\
\label{ImDyn} \Im
m\Pi_0^{ret}(q,\Omega,T)&=&\mbox{sgn}\Omega\frac{Nq^2}{4\pi\sqrt{\Omega^2-q^2}}
\int_{-1}^1dx\sqrt{1-x^2}\left[\frac{1}{2}-\frac{1}{1+\exp{\frac{|\Omega|+qx|}{2T}}}\right]
\end{eqnarray}
while for $\bq^2>\Omega^2$
\begin{eqnarray}
\label{ReStat} \Re e\Pi_0^{ret}(q,\Omega,T)&=&NT\frac{\ln2}{\pi}+
\frac{Nq^2}{4\pi\sqrt{q^2-\Omega^2}}\int_{-1}^{1}dx\sqrt{1-x^2}\left[\frac{1}{2}-\frac{1}{1+\exp\left|\frac{qx+|\Omega|}{2T}\right|}\right]\\
\label{ImStat} \Im
m\Pi_0^{ret}(q,\Omega,T)&=&\mbox{sgn}\Omega\frac{Nq^2}{4\pi\sqrt{q^2-\Omega^2}}\int_{1}^{\infty}dx
\left[\frac{\sqrt{1-x^2}}{1+\exp{\frac{qx-|\Omega|}{2T}}}-\frac{\sqrt{1-x^2}}{1+\exp{\frac{qx+|\Omega|}{2T}}}\right]
\end{eqnarray}
\end{widetext}
At $T=0$, $\Pi^{ret}_0(q,\Omega,0)=$
$$
\frac{N}{16}\frac{q^2}{\sqrt{|q^2-\Omega^2|}}\left\{\theta(q^2-\Omega^2)+i\mbox{sgn}\Omega\;\theta(\Omega^2-q^2)\right\}.
$$
Note that the interaction correction to the free energy has the
scaling form $$\delta
f(T,N,e_{\ast}^2,\Lambda)=T^3\mathcal{F}(e_{\ast}^2N,\Lambda/T),
$$
and in the low temperature limit of interest $\frac{\Lambda}{T}\gg
1$.
I now proceed with the evaluation of Eq.(\ref{deltaf}) in this
limit.

Consider first the static regime $q^2>\Omega^2$. As can be readily
seen from Eq.(\ref{ImStat}), for large $q$ $\Im
m\Pi_0^{ret}(q,\Omega,T)$ is small compared to $q/e_{\ast}^2$ and
the real part, and the latter can be approximated by $\Re
e\Pi_0^{ret}(q,\Omega,0)$ to the same order. As a result, the
contribution to $\delta f$ from the static regime is
\begin{eqnarray}
T^3\int^{\frac{\Lambda}{T}}\frac{dq
q^2}{2\pi}\frac{Ne_{\ast}^2}{4\pi}\int_0^1\frac{dy}{2\pi}\frac{1}{\sqrt{1-y^2}+\frac{Ne_{\ast}^2}{16}}\nonumber\times\\
\int^{\infty}_1dx\left[\frac{\sqrt{x^2-1}}{1+\exp\frac{q(x-y)}{2}+1}-\frac{\sqrt{x^2-1}}{1+\exp\frac{q(x+y)}{2}+1}\right]\nonumber\\
\rightarrow
\frac{8\zeta(\frac{5}{2})}{\pi^{5/2}}\left(1-\frac{\sqrt{2}}{4}\right)\sqrt{\frac{\Lambda}{T}}
-\frac{48\zeta(3)}{\pi^2}\frac{1}{Ne_{\ast}^2}\ln\frac{\Lambda}{2T}\nonumber
\end{eqnarray}
where the last line represents only the most divergent contribution.

Similarly, the most singular contribution from the dynamic regime,
$\Omega^2>q^2$, can be evaluated by expanding (\ref{deltaf}) to
first order in deviation of $\Im m\Pi^{ret}_0(q,\Omega,T)$ from $\Im
m\Pi^{ret}_0(q,\Omega,0)$ as well as to the first order in $\Re
e\Pi^{ret}_0(q,\Omega,T)$, both of which vanish at large $q$. The
algebra is somewhat tedious and I just state the final result for
the asymptotic expansion in $\Lambda/T$:
\begin{eqnarray}
&-&\frac{8\zeta(\frac{5}{2})}{\pi^{5/2}}\left(1-\frac{\sqrt{2}}{4}\right)\sqrt{\frac{\Lambda}{T}}-
\frac{48\zeta(3)}{\pi^2}\frac{1}{Ne_{\ast}^2}\ln\frac{\Lambda}{2T}\nonumber\\
&+&\frac{3\zeta(3)}{32\pi^3}N^2e_{\ast}^4\times
g\left(\frac{Ne_{\ast}^2}{16}\right)\times\ln\frac{\Lambda}{2T}\nonumber
\end{eqnarray}
where the function $g(x)$ is
\begin{eqnarray}
g(x)=\int_0^{\infty}dy\frac{(1+y)^2}{\sqrt{y(2+y)}(y(2+y)+x^2)^2}
\end{eqnarray}
This integral can be evaluated by a substitution followed by a
partial fraction decomposition and one finds
\begin{eqnarray}\label{g}
g(x)&=&\frac{1}{2x^2}+\frac{\tan^{-1}\left[\frac{\sqrt{1-x^2}}{x}\right]}{2x^3\sqrt{1-x^2}};\;\;\;\; 0<x<1\nonumber\\
g(x)&=&\frac{1}{2x^2}+\frac{1}{4x^3\sqrt{x^2-1}}\ln\left[\frac{x+\sqrt{x^2-1}}{x-\sqrt{x^2-1}}\right];\;\;\;\;
x>1\nonumber\\
\end{eqnarray}

Combining everything together I find that up to
$\mathcal{O}(\frac{1}{N})$:
\begin{eqnarray}\label{fLog}
f=-N\frac{3\zeta(3)}{4\pi}T^3+\left[\frac{24\zeta(3)}{\pi^3}\lambda^2g(\lambda)-\frac{6\zeta(3)}{\pi^2}\frac{1}{\lambda}\right]T^3\ln\frac{\Lambda}{2T}\nonumber\\
\end{eqnarray}
where $\lambda=Ne_{\ast}^2/16$. This is the result quoted in
Eq.(\ref{deltaf1}). The corresponding function $
\eta=\frac{1}{N}\left[\frac{16}{\pi^2}\lambda^2g(\lambda)-\frac{4}{\pi\lambda}\right]$
is plotted in Fig. (\ref{etaPlot}). Note that the contribution from
small $q$, which also includes the effect of the thermo-plasma mode
\cite{Vafek2006}, gives only terms of order $T^3$.

\begin{figure}[t]
\begin{center}
\includegraphics[width=0.41\textwidth]{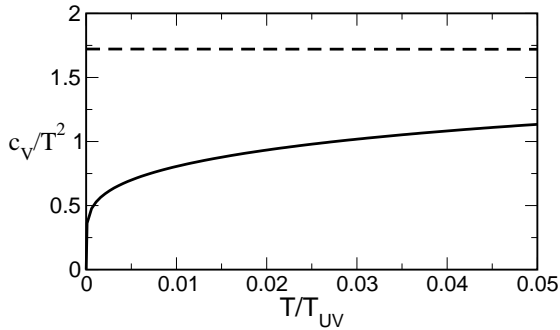}
\end{center}
\caption{The suppression of the specific heat (per Dirac cone)
relative to the non-interacting result
$c^{(0)}_V=\frac{9\zeta(3)}{2\pi}k_B^3T^2/(\hbar v_F)^2$ (dashed
line) vs. $T/T_{UV}$. This plot is for $N=4$, $v_F=10^6m/s$,
$\eps=1$ which gives $\eta\approx 0.14$.} \label{cVPlot}
\end{figure}

If we interpret the above expression as an expansion of the free
energy given in Eq.(\ref{fansatz}), the low temperature specific
heat is subquadratic. Fig. (\ref{cVPlot}) shows this suppression for
realistic material parameters pertinent to the single-layer
graphene.

It is interesting to make a connection between this result and a
seemingly unrelated one of Ref.\#\cite{KimLeeWen1997}, where the
authors studied the specific heat of $N$ massless 2D Dirac fermions
coupled only to the space component of a 2D gauge field within the
large $N$ expansion. Unlike our Eq.(\ref{deltaf1}), the correction
to the free energy found numerically in \cite{KimLeeWen1997} has the
opposite sign, i.e. the specific heat is enhanced relative to the
non-interacting case. The numerical coefficient of the
$T^2\ln\frac{1}{T}$ {\em enhancement} of $c_V$ is $2.79
$\cite{KimLeeWen1997}, while in this work I find that upon setting
$\lambda\rightarrow \infty$ the coefficient of the specific heat
$T^2\ln T$ {\em suppression} is $72\zeta(3)/\pi^3\approx 2.791$. The
opposite signs of these terms can be understood as follows: setting
$\lambda$ to infinity in Eq.\#(\ref{deltaf1}), recovers the leading
logarithmic contribution to the specific heat of QED$_3$ within the
large $N$ expansion arising from coupling solely to the {\em time}
component of the gauge field. However, due to the Lorenz invariance
of QED$_3$, the leading logarithmic contributions from the time and
the space components of the gauge field must cancel
exactly\cite{Vafek2003}. It is in this sense that the result in
Eq.(\ref{deltaf1}) is consistent with Ref.\cite{KimLeeWen1997}.

In conclusion, I found that the thermodynamic signature of the long
range electron-electron interactions in a two dimensional semimetal
are much more pronounced than in the case of a metal. This stems
from the inability of the charge carriers to screen the long range
interactions, and as a result the density of states is effectively
suppressed near the Dirac point. The experimental observation of the
effect would be an important step towards our understanding of
critical massless matter.


I wish to thank Professor Tesanovic for useful discussions.
\bibliography{th2}

\begin{thebibliography}{18}
\bibitem[{con()}]{cone}
\bibinfo{note}{Not counting spin, there are two modes associated with each
  Dirac cone.}
\bibitem[{log()}]{logs}
\bibinfo{note}{Showing that this dispersion indeed leads to $f\sim -T^3/\ln^2T$
  as $T\rightarrow 0$ is facilitated by using the (secondary branch of the)
  Lambert W function and its asymptotic expansion, c.f.\cite{Corless1996}}.
\end{thebibliography}

\end{document}